\RequirePackage[
datamodel=acmdatamodel,
style=acmauthoryear, 
]{biblatex}

\documentclass[acmsmall, screen, natbib=false, nonacm]{acmart}

\usepackage{ifthen}
\newboolean{showcomments}
\setboolean{showcomments}{true} 
\ifthenelse{\boolean{showcomments}}
  {\newcommand{\nb}[2]{
    \fcolorbox{gray}{yellow}{\bfseries\sffamily\scriptsize#1}
    {\sf\small$\blacktriangleright$\textit{#2}$\blacktriangleleft$}
    \addcontentsline{toc}{subsection}{\textit{\textbf{TODO:#2}}}
   }
   
  }
  {\newcommand{\nb}[2]{}
   
  }

\newcommand\supports[2]{#1\odot#2}

\newcounter{rqcounter}
\usepackage{cleveref}
\usepackage{subcaption}
\usepackage{pgfplots}
\usepackage{simplebnf}
\usepackage{currfile-abspath}
\getabspath{\jobname.log}
\usepackage[frozencache=true,cachedir=minted-cache]{minted}
\usepackage{tikz}
\usepackage{stackengine}
\usepackage{calc}
\usepackage{mathtools}
\usepackage{tikz-cd}
\usepackage{csquotes}
\usepackage{mathpartir}
\usepackage{circledsteps}

\usetikzlibrary{decorations.pathmorphing}
\usetikzlibrary{decorations.pathreplacing}
\usetikzlibrary{positioning}
\usepgfplotslibrary{fillbetween}
\usepgflibrary{patterns}
\usetikzlibrary{patterns}
\usetikzlibrary{tikzmark}
\usetikzlibrary{math}

\setcopyright{acmcopyright}
\copyrightyear{2023}
\acmYear{2023}
\acmDOI{XXXXXXX.XXXXXXX}

\acmConference[Conference acronym 'XX]{Make sure to enter the correct
  conference title from your rights confirmation email}{June 03--05,
  2018}{Woodstock, NY}
\acmISBN{978-1-4503-XXXX-X/18/06}

\pgfplotsset{compat=1.18}
\begin{document}

\title{Sound Interval-Based Synthesis for Probabilistic Programs}


\author{Guilherme Espada}
\email{gjespada@ciencias.ulisboa.pt}
\orcid{0000-0001-8128-7397}
\affiliation{%
  \institution{LASIGE, Faculdade de Ciências da Universidade de Lisboa}
  \city{Lisboa}
  \country{Portugal}
}

\author{Alcides Fonseca}
\email{alcides@ciencias.ulisboa.pt}
\orcid{0000-0002-0879-4015}
\affiliation{%
  \institution{LASIGE, Faculdade de Ciências da Universidade de Lisboa}
  \city{Lisboa}
  \country{Portugal}
}


\begin{abstract}

%
Probabilistic programming has become a standard practice to model stochastic events and learn about the behavior of nature in different scientific contexts, ranging from Genetics and Ecology to Linguistics and Psychology.
However, domain practitioners (such as biologists) also need to be experts in statistics in order to select which probabilistic model is suitable for a given particular problem, relying then on probabilistic inference engines such as Stan, Pyro or Edward to fine-tune the parameters of that particular model. Probabilistic Programming would be more useful if the model selection is made automatic, without requiring statistics expertise from the end user.

Automatically selecting the model is challenging because of the large search space of probabilistic programs needed to be explored, because the fact that most of that search space contains invalid programs, and because invalid programs may only be detected in some executions, due to its probabilistic nature.

We propose a type system to statically reject invalid probabilistic programs, a type-directed synthesis algorithm that guarantees that generated programs are type-safe by construction, and an heuristic search procedure to handle the vast search space.

We collect a number of probabilistic programs from the literature, and use them to compare our method with both a type-agnostic random search, and a data-guided method from the literature (DaPPer). Our results show that our technique both outperforms random search and DaPPer, specially on more complex programs.
This drastic performance difference in synthesis allows for fast sampling of programs and enables techniques that previously suffered from the complexity of synthesis, such as Genetic Programming, to be applied.

\end{abstract}



\maketitle

\newcommand{\EM}[1]{\ensuremath{#1}}
\newcommand{\ssymbol}[1]{\EM{#1}}
\newcommand{\bnfdef}{\EM{\vcentcolon\vcentcolon=}}
\newcommand{\emphbf}[1]{\textbf{\emph{#1}}}
\newcommand{\spmid}{\EM{\ \mid \ }}
\newcommand{\vsample}[1]{\EM{\mathit{sample}(#1)}}
\newcommand{\vconst}{\EM{\mathsf{c}}}
\newcommand{\anyval}{\ssymbol{v}}
\newcommand{\anydist}{\ssymbol{d}}
\newcommand{\dnormal}[2]{\EM{\mathit{Normal}(#1,#2)}}
\newcommand{\duniform}[2]{\EM{\mathit{Uniform}(#1,#2)}}
\newcommand{\lang}{\textit{PGPLang}}

\newcommand{\xleadsto}[2]{%
    \mathrel{%
        \begin{tikzpicture}[baseline=-.75ex]
        \node[%
            ,inner sep=.5ex
            ,align=center
            ] (tmp) {$\hphantom{\ensurestackMath{\stackengine{0pt}{\scriptstyle #2} {\scriptstyle #1}{O}{c}{F}{F}{L}}}$};
        \path[%
            ,draw,<-
            ,decorate,decoration={%
                ,zigzag
                ,amplitude=0.7pt
                ,segment length=1.2mm,pre length=3.5pt
                }
            ]
        (tmp.east) -- node [midway, below,align=center] {$\scriptstyle #2$} node [midway, above, align=center] {$\scriptstyle #1$}  (tmp.west);
        \end{tikzpicture}
        }
    }
\section{Introduction}


Probabilistic programming is a paradigm that treats distributions and uncertainty as first class citizens. Distributions can be combined, composed, and manipulated similar to functions in a functional programming language.

\pgfmathdeclarefunction{gauss}{2}{%
  \pgfmathparse{1/(#2*sqrt(2*pi))*exp(-((x-#1)^2)/(2*#2^2))}%
}
\pgfmathdeclarefunction{gaussnoisy}{2}{%
  \pgfmathparse{(1/(#2*sqrt(2*pi))*exp(-((x-#1)^2)/(2*#2^2)))*(1+((sin(2*x*200)+sin(pi*x*200))*0.05))}%
}

\begin{figure*}
  \centering
  \begin{tikzpicture}[minimum size = 2em]

  \def\startx{-5}    
  \def\endx{5}       
  \def\camean{-1.0}  
  \def\casigma{1.0}  
  \def\cbmean{1.0}   
  \def\cbsigma{1.0}  
  \def\verticala{0.7} 
  \def\verticalb{1.3} 
  \node[at={(2cm,1.2cm)}, anchor=mid] (plabel) {$\mathit{prior}$};
  \begin{axis}[
  at={(0,0)},
  domain=\startx:\endx,
  samples=101,
  ymax=0.5,
  enlargelimits=false,
  axis x line=middle,
  axis y line=middle,
  xtick={\empty},
  ytick={\empty},
  height=2.5cm,
  width=5cm
  ]
  \addplot [thin, smooth] {gauss(\camean,\casigma)};

  \end{axis}

  \node[at={(4.5cm,1.2cm)}, anchor=mid] {$\mathit{conditioned\ on}$};

  \node[at={(7cm,1.2cm)}, anchor=mid] {$\mathit{evidence}$};
  \begin{axis}[
      at={(5cm,0)},
      domain=\startx:\endx,
      samples=101,
      ymax=0.5,
      enlargelimits=false,
      axis x line=middle,
      axis y line=middle,
      xtick={\empty},
      ytick={\empty},
      height=2.5cm,
      width=5cm
      ]
      \addplot [name path=f, thin, smooth, opacity=0.2] {gaussnoisy(\cbmean,\casigma)};
      \path [name path=xaxis] (\pgfkeysvalueof{/pgfplots/xmin},0) -- (\pgfkeysvalueof{/pgfplots/xmax},0);
      \addplot[red!5, opacity=1,pattern=grid, pattern color=red!90] fill between [of=f and xaxis];

      \end{axis}
      \node[at={(9.5cm,1.2cm)}, anchor=mid] {$=$};
      \node[at={(12cm,1.2cm)}, anchor=mid] {$\mathit{posterior}$};
  \begin{axis}[
      at={(10cm,0)},
      domain=\startx:\endx,
      samples=101,
      ymax=0.5,
      enlargelimits=false,
      axis x line=middle,
      axis y line=middle,
      xtick={\empty},
      ytick={\empty},
      height=2.5cm,
      width=5cm
      ]
      \addplot [thin, smooth] {gauss(0,\casigma)};

      \end{axis}

  \end{tikzpicture}
  \caption{Diagram demonstrating Bayesian Inference. Given some \textit{prior} (assumptions about the system) and \textit{evidence} (observations) a \textit{posterior} distribution is produced, which better matches the evidence. Commonly, multiple steps of this process are iterated, where the \textit{posterior} of one iteration is used as the \textit{prior} of the next.}
  \label{fig:gdiagram}
\end{figure*}
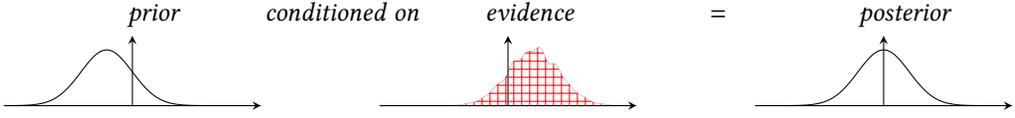

Probabilistic Programming and Bayesian Methods have seen popularity as a way to model stochastic or uncertain events in various fields, such as Computer Vision~\cite{3DP3}, Physics~\cite{physics1}, Biology~\cite{health1}, Pharmacology~\cite{pharm1}, Epidemiology~\cite{epidem1} Chemistry~\cite{chem1}, and Economics~\cite{econ1}, among others. Probabilistic Programming frameworks often provide tools not only to build these programs, but also to fine-tune them with techniques such as Markov Chain Monte Carlo~\cite{mcmc} or Stochastic Variational Inference~\cite{svi} to transform the models so that they more closely representing actual observed data.

These models can then be used in a variety of tasks. Examples include:
\begin{itemize}
  \item Predicting future events based on past observations, such as disease spread~\cite{epidem1}.

  \item Estimating the probability of certain outcomes given observed data, such as the influence of certain mutations on cancer~\cite{health1}.

  \item Explaining the stochastic process by which as event occurs, such as the choices of external economic actors~\cite{econ1}.
\end{itemize}

Let us consider the example in which, due to some prior knowledge or assumption, it is thought that some phenomenon follows a normal distribution centered on $-1$. This knowledge forms the \textit{prior} distribution. We can encode it for example, as $\mathit{Normal}(-1, 1)$. Nonetheless, it is not common that the \textit{prior} is the best model of the phenomena. If this was always the case, Bayesian methods would not be needed. Now, say we have measured some real-world occurrences of this phenomenon we are trying to model. These observations form our \textit{evidence}. Given these observations, we can now update the \textit{prior} to better match the \textit{evidence}, producing a new model, the \textit{posterior}, by adjusting the parameters of the distribution. This workflow is summarized in \cref{fig:gdiagram}.

While we chose a simple \textit{prior} in our example, in realistic models, the complexity can increase drastically by having parameters to one distribution that are sampled from other distributions. Furthermore, a good \textit{prior} (i.e., one that matches the probabilistic structure) is essential for the algorithms used to converge and produce a good \textit{posterior}.

Designing these \textit{prior}s requires a domain expert and their insights on the phenomena that is being modelled. Additionally, to reify these assumptions into a good \textit{prior} an expert in Bayesian methods is also essential. This modeling is deep and complex enough that several books are dedicated to this matter~\cite{statrethink,probmodels}.

It is difficult for a non-expert user to use these techniques. It may be the case that an expert in the domain under analysis is not available, that they are not versed in Bayesian modeling, or even that the domain is novel enough that an expert does not exist. Therefore, it is desirable to be able to create a model from only the \textit{evidence}, without requiring application of domain knowledge to build a \textit{prior}.

There are some techniques such as Mixtures of Gaussians~\cite{mixofgaussians}, Variational Autoencoders~\cite{vae}, or Probabilistic Neural Networks~\cite{probnn} that are able to model stochastic events without the need to specify a prior.

Mixture of gaussians approximate arbitrary distributions as sums of gaussians. This results in any non-gaussian distribution necessarily requiring multiple terms to obtain a reasonable approximation. Similarly, Variational Autoencoders and Probabilistic Neural Networks, being Neural Network based techniques, typically use multiple layers, again resulting in expressions with numerous terms.

Thus, these have a key shortcoming: the resulting models are large and thus expensive to evaluate and difficult to interpret.

It is clear why a small model can be useful: it takes less resources to store and evaluate. However, it might be less clear why an \textit{interpretable} model is desirable. If the model is interpretable enough, then the model can act as a \textit{distillation} of the information present in the raw data (as explored in \citeauthor{baysynth}).

Program synthesis systems can be tuned to produce models that are both small and interpretable (by, for example, limiting the size of output expressions).

However, synthesis of probabilistic programs poses unique challenges. Given a probabilistic program, expressions that may produce faulting evaluations may not be discovered, even in multiple runs\footnote{Given infinite runs, it will \textit{almost surely} be discovered, however, it is still not guaranteed.}. \Cref{rareinvalid} shows an example of this phenomenon. An invalid execution occurs with a very low probability, making it difficult to dynamically validate.

Intuitively, a model that may fault, provides less insight that a model that cannot. In addition, models that fault can prove challenging to compose, due to the error rates of each individual component composing\footnote{A model M with two components A and B, each with an individual failure rate of $a$ and $b$ has a composed failure rate of $1-((1-a)(1-b)).$}.

\begin{figure}
  \centering
  \begin{align*}
  x &\sim \overbrace{\textit{Normal}(10, 1)}^{\mathbb{P}(x \le 0) \approx 1^{-23}} \\
  \textit{obs} &\sim \textit{Normal}(10, x)
  \end{align*}
  \caption{An invalid mixture distribution. The second parameter of the \textit{Normal} distribution must be positive. However, $x$ is not guaranteed to be positive. This distribution has a low chance of being detected as invalid by sampling, because the invalid parameter is very unlikely to be produced.}
  \label{rareinvalid}
\end{figure}

Additionally, some common distributions' support depends on their exact parameterization, making them effectively dependently typed. An example of this is the \textit{Uniform} distribution, whose support is the range between its two parameters. Therefore, if its parameters are random variables, it is generally not possible to statically determine the exact support.

\begin{figure}
  \centering
  \includegraphics[width=0.5\linewidth]{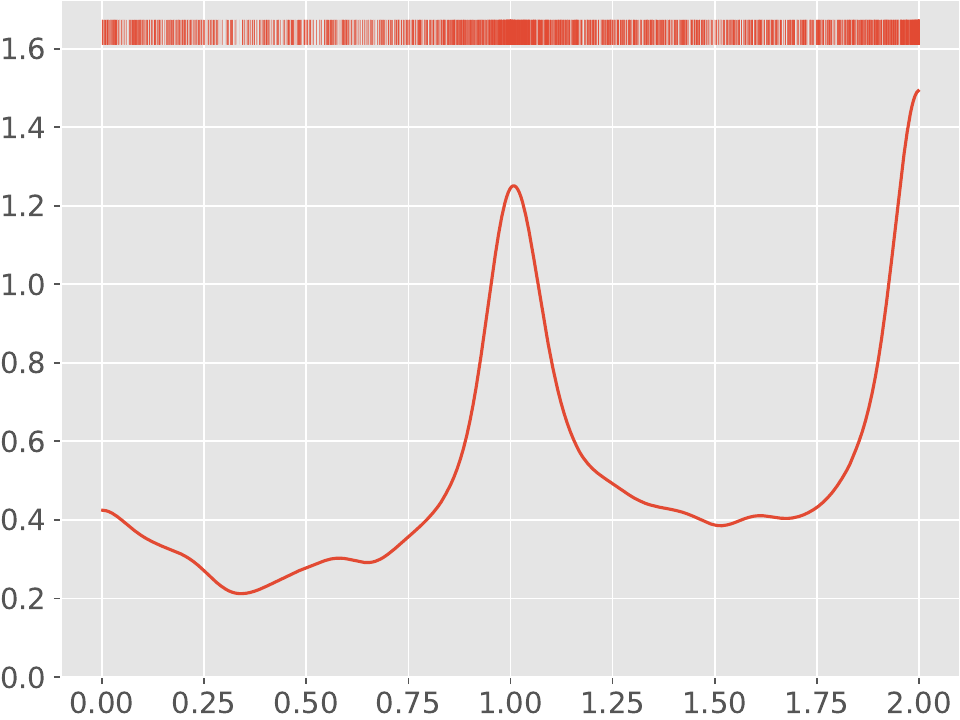}\\
  $\downarrow$\\
  $\textit{add}(\textit{Beta}(0.3, 0.25), \textit{Beta}(0.4, 0.25))$
  \caption{Example of a Stochastic Symbolic Regression process input and output. In the top of the figure, each vertical line represents a discrete sample. Under it, the plotted line represents the kernel-smoothed empirical probability density function. The expression shows the desired result from the regression, which corresponds to the function that generated the initial samples.}
  \label{fig:problemexample}
\end{figure}

\paragraph{Problem Statement} Given some \textit{evidence} and some time or size budget, synthesize the probabilistic program that best describes the given \textit{evidence}, without needing to specify a \textit{prior}. An example of such a problem is shown in \cref{fig:problemexample}.

One common technique for modeling from data is Symbolic Regression.
Symbolic Regression is a family of machine learning techniques wherein a mathematical closed-form expression is extracted from samples of known data, without requiring domain knowledge. If we consider that closed-form mathematical expressions are programs, Symbolic Regression can be seen as a subdomain of example-driven synthesis.

However, as previously noted, in the Probabilistic Programming domain, a large part of the search space consists of invalid programs. In addition, it is not possible to verify correctness via execution.

To ensure run-time correctness of Probabilistic Programs, we propose a type system that tracks the possible values that Random Variables can take. We then build a synthesis procedure on top of this type system, to enable efficient top-down synthesis of probabilistic programs.

We apply this procedure to a small probabilistic programming language, \textit{\lang{}}.
While \lang{} is quite simple, it captures the key challenges in synthesizing probabilistic programs. The language includes four primitive distributions: Normal, Uniform, Laplace, and Beta. These distributions were chosen as they are commonly used in probabilistic modeling and cover a range of different behaviors (for example, bounded vs. unbounded support). The language also includes the addition operator, allowing for more complex distributions to be built up from the primitives.

\paragraph{Contributions}
In summary, this work provides the following contributions:
\begin{itemize}
  \item \textit{\lang{}}, a small probabilistic programming language
  \item An interval-based dependent typing strategy for probabilistic programs
  \item A synthesis algorithm that is sound with regard to the type system.
  \item A comparison, in a stochastic symbolic regression-like setting, of the algorithm in contrast to both untyped synthesis and a data-guided synthesis procedure~\cite{blogsynthesis}.
\end{itemize}

\section{Approach}
As previously mentioned, the main challenge with synthesizing probabilistic programs is preventing the generation of invalid programs. While in traditional, deterministic program synthesis, the execution of the result can be used to discard invalid programs, some probabilistic programs only rarely show evidence of being invalid, making the process of invalidating programs through execution impossible. Thus, sampling and testing programs from a context-free grammar does not guarantee that only valid probabilistic programs are generated. This may lead to problems for the user later in the modeling process.

Our approach consists of a synthesis algorithm that tracks the possible ranges of random variables in a top-down approach. 
In particular, we propose a type system to distinguish valid probabilistic programs from invalid ones. Our synthesis algorithm is guaranteed to generate valid programs with respect to this type system. We then show soundness of this algorithm.

To give context to this development, we first present the \lang{} probabilistic programming language, that includes key concepts present in many real-world probabilistic languages.

%
%

\subsection{The \lang{} Language}
\lang{} was designed to be a trade-off between being simple enough to concisely explain our approach, but also representing the challenges of real-world probabilistic languages. More precisely, it only supports a limited number of distributions, it only has a single deterministic operator and does not support control-flow. Other distributions can be easily added to the language, requiring only the proper typing rule to be deduced from the support. Deterministic operators can also be introduced, as long as the impact on the values can be reasoned about%
. Finally, control-flow mechanisms can be introduced by conservatively merging the possible values from both branches, as it is common in symbolic execution. Alternatively, control-flow can be propagated to the top-level of the program, which represents multiple simple type-checking/synthesis problems, similarly to how it is handled by DaPPer~\cite{blogsynthesis}.


\begin{figure}
\begin{bnfgrammar}
expr : Expressions ::= val
    | \underline{let} val \underline{in} expr
;;
val : Values ::= distr
    | op
;;
distr : Distributions ::= \underline{Normal} arg arg
    | \underline{Uniform} arg arg
    | \underline{Laplace} arg arg
    | \underline{Beta} arg arg
;;
op : Deterministic Operators ::= \underline{add} arg arg
;;
arg : Function Arguments ::= var
    | lit
;;
var : Bound References ::= \underline{v} \textit{\{int\}}
;;
lit : Literal Values ::= \textit{\{float\}}
\end{bnfgrammar}
\caption{The grammar of \lang{}. Literal tokens are \underline{underlined}. Productions in \{braces\} are well-known, and have been omitted for brevity.}
\label{fig:grammar}
\end{figure}

The grammar of \lang{} is described in \cref{fig:grammar}. The language supports floating-point literals (\textit{lit}), variables (\textit{var}), distributions (\textit{distr}, modelled as functions), the deterministic operator \textit{add}, and let expressions.

\newcommand{\graphicmath}[1]{%
 {\mathchoice%
  {\includegraphics[height=1.6ex]{#1}}%
  {\includegraphics[height=1.6ex]{#1}}%
  {\includegraphics[height=1.2ex]{#1}}%
  {\includegraphics[height=0.9ex]{#1}}%
 }%
}
\begin{figure}
\centering%
\begin{minipage}{0.5\textwidth}%
\begin{minted}[linenos, texcomments]{haskell}
let Beta 0.3 0.25 in -- $\Gamma = [\graphicmath{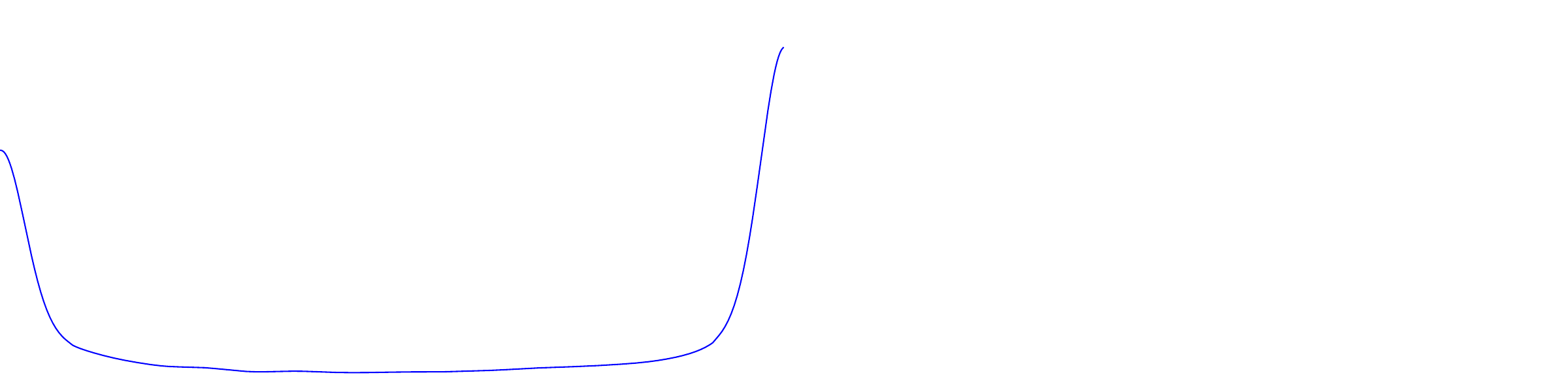}]$
let Beta 0.4 0.25 in -- $\Gamma = [\graphicmath{figure_scripts/examplerefs-l1.pdf}, \graphicmath{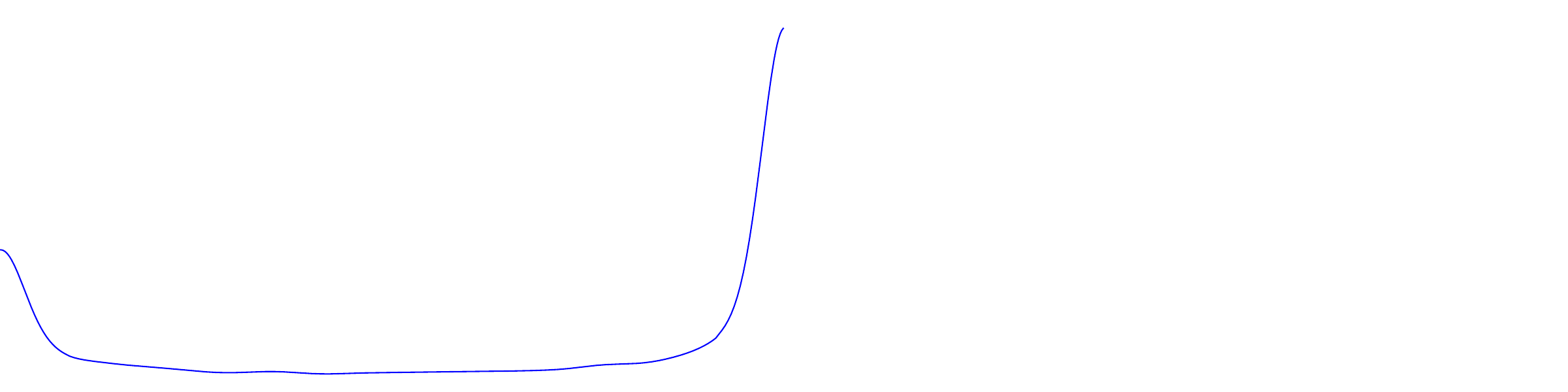}]$
add v1 v2            -- Output: $\graphicmath{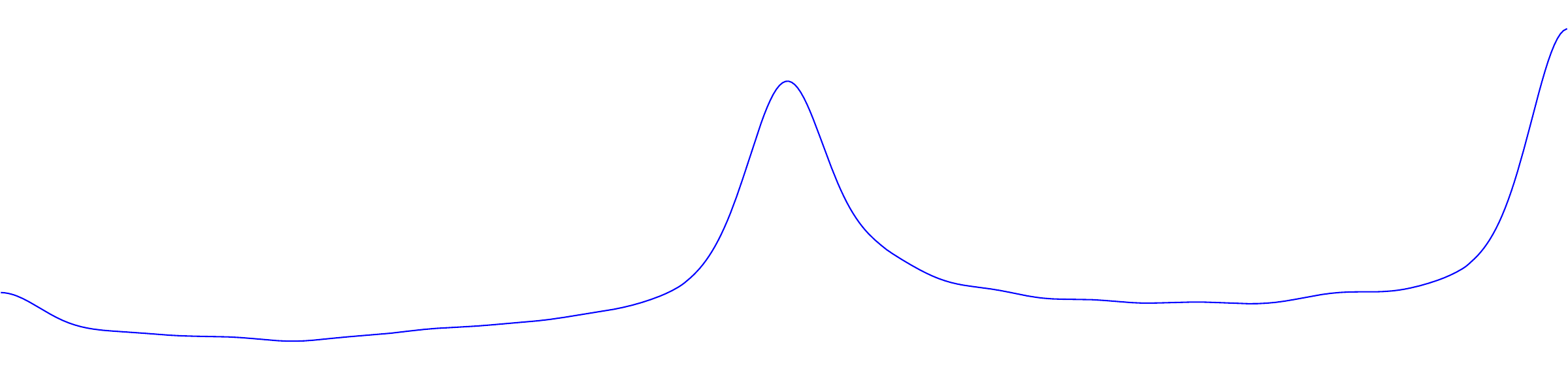}$
\end{minted}
\end{minipage}
\caption{Evolution of bound references during program evaluation. We display a graphical representation of the distribution, with the end result being the same as in \cref{fig:problemexample}.}
\label{fig:bindingdiagram}
\end{figure}

To ease the tracking of types, the design of the grammar imposes the A-normal Form (ANF): nested operators are not allowed, requiring intermediate values to be bound using the \textit{let} expression.
As let expressions are nameless, bound values are pushed into the bottom of the context and indexed by their positions from the top, resembling De Bruijn levels. The code examples shown in this paper are formatted such that the binding of $v{\color{red}n}$ always occurs on line number~${\color{red}n}$. An example of the evolution of the context during execution is shown in \cref{fig:bindingdiagram}.

For illustrative purposes we'll take as an example the following situation:

\begin{quote}
  I will first randomly (and uniformly) pick a real number between 0 and 1. We'll call this the target. I will then throw a dart at the target. As my aim is not perfect, the number I actually hit might be different from my target. We will assume it follows a normal distribution centered on my target, with a deviation of 0.1.
\end{quote}

We are interested in describing the distribution of the number hit by my dart. In Python, this could be written as \mintinline{python}{return gauss(uniform(0, 1), 0.1)}, similarly to other probabilistic programming languages. In our language, A-Normal form is enforced, resulting in the equivalent program \mintinline{haskell}{let uniform 0 1 in Normal v1 0.1}.



\subsection{Type system}

We propose a novel type system to identify the subset of programs that will always be sampleable. The main idea is to track the under- and overapproximations of the values throughout the program. For this, our type system is based on interval arithmetic.

We consider only the single type family of stochastic values. Deterministic values can be modelled as a stochastic value with a very narrow bound that only allows that deterministic value to be sampled.

\begin{figure}
  \centering
  \begin{gather*}
      \frac{
              x_1 : \mathbb{R}
              \quad
              x_2 : \mathbb{R}
              \quad
              x_1 \le x_2
          }{
              \underbrace{
                  [x_1, x_2]
              }_{
                  = [x]
              } : \mathit{Bound}
      }
      \tag{Form-CBound}
      \label{eqn:fcbound}
      \\
      \frac{}{
              \underbrace{
                  \emptyset
              }_{
                \text{\clap{the empty bound is a valid bound}}
              } : \mathit{Bound}
      }
      \tag{Form-EBound}
      \label{eqn:febound}
      \\
      \frac{
          \overbrace{
              [d^1] : \mathit{Bound}
          }^\text{
              underapproximation
          }
          \quad
          \overbrace{
              [d^2] : \mathit{Bound}
          }^\text{
              overapproximation
          }
          \quad
          [d^1] \subseteq [d^2]
      }{
          \underbrace{
              {\text{«}}[d^1], [d^2]{\text{»}}
          }_{
              = {\text{«}}d{\text{»}}
          } : \mathit{Type}
      }
      \tag{Form-DualBound}
      \label{eqn:formtype}
\end{gather*}
\caption{Type Formation Rules}
\label{fig:formation}
\end{figure}

We denote our types by \({\text{«}}[d^1], [d^2]{\text{»}}\) (\cref{eqn:formtype}) where the two intervals inside the double bound correspond to the underapproximation and the overapproximation, respectively, of the possible values of that typed expression. We write \({\text{«}}d{\text{»}}\), as a shorthand for \({\text{«}}[d^1], [d^2]{\text{»}}\). \Cref{fig:formation} describes the type formation rules of the language, namely that the underapproximation needs to be contained in the overapproximation, and that each interval can be empty (\cref{eqn:febound}) or not (\cref{eqn:fcbound}).

Because we want to relate the bounds of values with the support of distributions, we use the following notation: we write \(\supports{dist}{x}\) to mean that \(dist\) \emph{supports} \(x\), or, in other words the probability of \(dist\) generating \(x\) is nonzero. We write \(sup(dist)\) to denote the set that is supported by the distribution, as given by \(sup(dist)=\{x:\supports{dist}{x}\}\).
Note that it is not necessary for every member of a family of distributions to have the same support. For example, \(\mathit{Uniform}(0,1)\) and \(\mathit{Uniform}(2,3)\) have the completely disjoint supports of \([0,1]\) and  \([2,3]\), respectively.

\begin{figure}
\begin{equation*}
  \frac{
    \overbrace{\mathit{dist}}^{\text{{any distribution}}} \quad
    \overbrace{{\text{«}}d{\text{»}}}^{\text{{any bound}}} \quad
    \forall x, x \in [d^1] \rightarrow \supports{\mathit{dist}}{x} \quad
    \forall x, {\supports{\mathit{dist}}{x} \rightarrow x \in [d^2]}
}
{
  \Gamma \vdash \textit{dist} : {\text{«}}d{\text{»}}
}
\end{equation*}
\caption{Typing Invariant}
\label{fig:tyinvariant}
\end{figure}


Our type system is based on the following important invariant, formalized in \cref{fig:tyinvariant}: A value from a distribution \(dist\) has type \({{{\text{«}}}[d^1], [d^2]{{\text{»}}}}\) if and only if it can generate (with nonzero probability) all elements of \([d^1]\) and can never generate elements outside \([d^2]\).

\begin{figure}
    \centering
    \begin{gather*}
        \frac{
                v : \mathbb{R}
            }{
                \Gamma \vdash v :  \underbrace{
                  {\text{«}}[v, v], [v, v]{\text{»}}
                }_{
                    \text{\clap{the only possible value a constant can have is itself}}
                }
        }
        \tag{Type-Literal}
        \\
        \frac{
          \Gamma \vdash \mu : {\text{«}}\emptyset, [-\infty, +\infty]{\text{»}} \quad
          \Gamma \vdash \sigma : {\text{«}}\emptyset, [0, +\infty]{\text{»}}
            }{
                \Gamma \vdash \mathit{Normal}(\mu, \sigma) : {\text{«}}[-\infty, +\infty], [-\infty, +\infty]{\text{»}}
        }
        \tag{Type-Normal}
        \label{eqn:t-normal}
        \\
        \frac{
          \Gamma \vdash a : {\text{«}}[a^1], [a^2]{\text{»}} \quad
          \Gamma \vdash b : {\text{«}}[b^1], [b^2]{\text{»}}
            }{
                \Gamma \vdash \mathit{Uniform}(a, b) : {\text{«}}\underbrace{a^1 \cap b^1, a^2 \cup b^2}_{
                  \mathclap{
                    \tiny\begin{split}
                      [x] \cap [y] &= [max(x_1, y_1), min(x_2, y_2)] \\
                      [x] \cup [y] &= [min(x_1, y_1), max(x_2, y_2)]
                    \end{split}
                  }
              }{\text{»}}
        }
        \tag{Type-Uniform}
        \\
        \frac{
          \Gamma \vdash \mu : {\text{«}}\emptyset, [-\infty, +\infty]{\text{»}} \quad
          \Gamma \vdash b : {\text{«}}\emptyset, [0, +\infty]{\text{»}}
            }{
                  \Gamma \vdash \mathit{Laplace}(\mu, b) : {\text{«}}[-\infty, +\infty], [-\infty, +\infty]{\text{»}}
        }
        \tag{Type-Laplace}
        \\
        \frac{
          \Gamma \vdash \alpha : {\text{«}}\emptyset, [0, +\infty]{\text{»}} \quad
          \Gamma \vdash \beta : {\text{«}}\emptyset, [0, +\infty]{\text{»}}
            }{
                  \Gamma \vdash \mathit{Beta}(\alpha, \beta) : {\text{«}}[0,1], [0,1]{\text{»}}
        }
        \tag{Type-Beta}
        \\
        \frac{
              \Gamma \vdash lv : lb \quad
              \Gamma \vdash rv : rb \quad
            }{
                \Gamma \vdash lv + rv : \underbrace{
                  \mathit{addDBounds}(lb, rb)
                }_{
                    \text{\clap{described in \cref{fig:adddbounds}}}
                }
        }
        \tag{Type-Add}
        \\
        \frac{
              \Gamma \vdash v : vb \quad
              \overbrace{\Gamma , vb}^{\text{\clap{append to context}}} \vdash body : bodyb \quad
            }{
                \Gamma \vdash \text{let}\ v\ \text{in}\ body : bodyb
        }
        \tag{Type-Let}
        \\
        \frac{
              \Gamma \vdash \overbrace{\Gamma({\color{red}n})}^{\text{\clap{index into context}}} : b
            }{
              \Gamma \vdash \text{v}\mathit{\color{red}n} : b
        }
        \tag{Type-Var}
        \\
        \frac{
            \Gamma \vdash v : {\text{«}}d{\text{»}} \quad
            [d^1] \supseteq [d'^1] \quad
            [d^2] \subseteq [d'^2] \quad
        }
        {
          \Gamma \vdash v : {\text{«}}d'{\text{»}}
        }
        \tag{Type-Subtyping}
        \label{eqn:dtyping}
  \end{gather*}
  \caption{Typing Rules}
  \label{fig:typing}
\end{figure}

\begin{figure}
\begin{equation*}
\begin{split}
	\mathit{addDBounds}({\text{«}}u,o{\text{»}},{\text{«}}u',o'{\text{»}})  =&\ {\text{«}}\mathit{addBounds}(u, u'), \mathit{addBounds}(o, o'){\text{»}} \\
	& \text{where} \\
	\mathit{addBounds}(\emptyset, [x]) =&\ \emptyset \\
	\mathit{addBounds}([x], \emptyset) =&\ \emptyset \\
	\mathit{addBounds}([x_1, x_2], [y_1, y_2]) =&\ [x_1+y_1, x_2+y_2]
\end{split}
  \end{equation*}
  \caption{Definition of addDBounds, merging two types using interval arithmetic.}
  \label{fig:adddbounds}
\end{figure}

To comply with this invariant, we developed the type system that we present in \cref{fig:typing}. We consider the judgement $\Gamma \vdash v : {\text{«}}d{\text{»}}$, stating that from context $\Gamma$, expression $v$ has type ${\text{«}}d{\text{»}}$. As mentioned before, literals are handled as having a singleton bound (Type-Literal). Different distributions have different requirements on their arguments, and present different images. To add a new distribution to the language, one would only need to identify these bounds. Let expressions (Type-Let) and variables (Type-Var) have traditional forms, assuming accesses are by indices from the beginning of the context. The deterministic $+$ operator takes the bounds of operands, and combines them following interval arithmetic (Type-Literal). We use the \texttt{addDBounds : ${\text{«}}d{\text{»}} \rightarrow {\text{«}}d{\text{»}} \rightarrow {\text{«}}d{\text{»}}$} function, defined in \cref{fig:adddbounds}, to apply the operator to the extremes of bounds.

Finally, our type system supports subtyping in cases where the given overapproximation bound is contained in the expected overapproximation, and the given underapproximation contains the expected underapproximation (Type-Subtyping).

This type system can be used to detect problems in existing probabilistic programs, but is also the basis to construct the synthesis algorithm in a type-safe way.

\subsection{Synthesis}

\newcommand{\synths}[5]{\EM{#1 \xleadsto{#3}{#4} \frac{#2}{#5}}}

\newcommand{\splitbudget}[3]{#2 \overset{#1}{\curlywedge} #3 }

As previously mentioned, synthesizing valid probabilistic programs is non-trivial. Firstly, because it is not possible to ensure validity by execution, and secondly because picking the wrong parameterization for distributions with dependent supports might require backtracking. However, now that we defined a type system for probabilistic programs, we can leverage it to build a correct synthesis algorithm. Furthermore, careful analysis of the supports of distributions with dependent types allows the algorithm to never end up in a dead-end situation. In this manner, we can build valid programs in a single pass, avoid both the need to backtrack, and the need to check the validity.

Our synthesis procedure takes as input a (possibly empty) context, a budget, and a target bound, and outputs a new context alongside a reference to the context cell containing the fresh expression. We write $\synths{\Delta}{\Delta'}{{\text{«}}t{\text{»}}}{b}{e}$ to mean that \textit{under context $\Delta$, an expression $e$ of type ${\text{«}}t{\text{»}}$ with a budget of $b$ is synthesized, returning a modified context, $\Delta'$}. \Cref{fig:synthesis} defines the synthesis rules.


\begin{figure}
  \centering
  \begin{gather*}
    \frac{
      \splitbudget{b}{b_\mu}{b_\sigma} \quad
      \synths{\Delta_i}{\Delta_\mu}{{\text{«}}[-\infty, +\infty], \emptyset{\text{»}}}{b_\mu}{\mu} \quad
      \synths{\Delta_\mu}{\Delta_\sigma}{{\text{«}}[0, +\infty], \emptyset{\text{»}}}{b_\sigma}{\sigma} \quad
        }{
            \synths{\Delta_i}{\Delta_\sigma}{{\text{«}}[-\infty, +\infty], [-\infty, +\infty]{\text{»}}}{b}{Normal(\mu, \sigma)}
    }
    \tag{Synth-Normal}
    \label{eqn:s-normal}
    \\
    \frac{
          }{
              \synths{\Delta}{\Delta'}{{\text{«}}[a, a], \_{\text{»}}}{0}{a}
      }
    \tag{Synth-Const-Single}
    \\
      \frac{
          r \sim \mathit{Uniform}(a, b)
          }{
              \synths{\Delta}{\Delta'}{{\text{«}}\emptyset, [a, b]{\text{»}}}{0}{r}
      }
    \tag{Synth-Const-Range}
    \\
    \frac{
      \splitbudget{b}{b_l}{b_r} \quad
      \synths{\Delta}{\Delta_l}{{\text{«}}\emptyset, [ol, ul]{\text{»}}}{b_l}{l} \quad
      \synths{\Delta_l}{\Delta_r}{{\text{«}}\emptyset, [uh, oh]{\text{»}}}{b_r}{r} \quad
          }{
              \synths{\Delta}{\Delta_r}{{\text{«}}[ul, uh],[ol, oh]{\text{»}}}{b}{\mathit{Uniform}(l, r)}
      }
    \tag{Synth-Uniform-Strict}
    \\
    \frac{
      \splitbudget{b}{b_l}{b_r} \quad
      s \sim \mathit{pick} [ol, oh]\quad
      \synths{\Delta}{\Delta_l}{{\text{«}}\emptyset, [ol, s]{\text{»}}}{b_l}{l} \quad
      \synths{\Delta_l}{\Delta_r}{{\text{«}}\emptyset, [s, oh]{\text{»}}}{b_r}{r} \quad
          }{
              \synths{\Delta}{\Delta_r}{{\text{«}}\emptyset,[ol, oh]{\text{»}}}{b}{\mathit{Uniform}(l, r)}
      }
    \tag{Synth-Uniform-Loose}
    \\
    \frac{
      \splitbudget{b}{b_l}{b_r} \quad
      {\text{«}}t_l{\text{»}}, {\text{«}}t_r{\text{»}} \sim \mathit{partition\_add}({\text{«}}t{\text{»}}) \quad
      \synths{\Delta}{\Delta_l}{{\text{«}}t_l{\text{»}}}{b_l}{l} \quad
      \synths{\Delta_l}{\Delta_r}{{\text{«}}t_r{\text{»}}}{b_r}{r} \quad
          }{
              \synths{\Delta}{\Delta_r}{{\text{«}}t{\text{»}}}{b}{\mathit{add}(l, r)}
      }
    \tag{Synth-Add}
\end{gather*}
\caption{Synthesis Rules}
\label{fig:synthesis}
\end{figure}

\begin{figure}
  \begin{equation*}
    \frac{
      b>0 \quad
      b_l \sim \mathit{Uniform}(0, \overbrace{b-1}^{\text{\clap{one unit of budget is consumed for the node itself}}}) \quad
      b_r = b - b_l
  }
  {
    \splitbudget{b}{b_l}{b_r}
  }
  \end{equation*}%
  \caption{Budget Splitting rule}
  \label{fig:splitbudget}
\end{figure}

During synthesis, a max budget is used. This represents the total amount of non-trivial (i.e., nor constants or binders) expressions the program can have. Non-terminal nodes consume one unit of budget, and must allocate the rest to its children. We chose to split the budget based on a uniform distribution, as shown in \cref{fig:splitbudget}.

Not all operators or distributions can be synthesized in the same manner, due to their domains and images.

For distribution families where the domain does not depend on the parameterization, the synthesis rules follow directly from the type checking rules. An example of this is shown in \cref{eqn:s-normal} (compare with \cref{eqn:t-normal}).

However, for distribution families where the domain is dependent on the parameterization, further care is needed. Taking as an example the $\mathit{Uniform}$ distribution, it is possible for a $\mathit{Uniform}$ with type ${\text{«}}[0,1], [0,1]{\text{»}}$ to type its arguments with either $({\text{«}}[0, 0], [0, 0]{\text{»}},\allowbreak {\text{«}}[0, 1], [0, 1]{\text{»}})$ or $({\text{«}}[0, 0.5], [0,0.5]{\text{»}},\allowbreak {\text{«}}[0, 0.5], [0, 0.5]{\text{»}})$, among other combinations. Thus, during synthesis, we must decide how to break up the bound. Here we follow a similar strategy to splitting the budget: We randomly pick a breakpoint within the bounds, and split the bound according to that point.

Let us now look at the $+$ operator. First, we'll invalidate the obvious solution: synthesizing the left side, and typing the right side to compensate. While this works for scalars, this does not work for intervals. This is due to the fact that in $a + b=c$, $b$ is not solvable for all fixed values of $a$ and $c$, unlike in the scalar domain\footnote{One possible intuition around this fact is that lengths are always positive, and summing intervals also sums their lengths. As such, if, in the equation $a + b=c$, you have an interval $a=[0,2]$ and $c=[0,1]$ then $b$ would have to have negative length.}.

However, if we consider the lengths of the intervals, it is also true that $|a| + |x| = |c|$. And since lengths are scalar, we can uniformly split them.

Intervals, however, cannot be unambiguously described by just their length, a start point is also needed. The strategy here is to fix one of $a$'s start points to the same as $c$, and calculate the other from the length. $b$'s points can then be calculated from $a$ and $c$. We then perturb both bounds by a random quantity in opposite directions. This guarantees that all possible combinations of valid types are generated.

One small detail is that the underapproximation must be contained within the overapproximation. As such, we transform our bounds into three intervals. If we have a bound ${\text{«}}[ul,uh], [ol, oh]{\text{»}}$ we transform it into the intervals $[ol, ul]$, $[ul, uh]$ and $[uh, oh]$. Once we have these bounds, we can then apply the previously described technique to each of them, in turn.

The final algorithm is somewhat more complex than described\footnote{We avoid any operation that may result in NaNs. For example, we avoid division of values that may be infinite, such as bound boundaries or bounds lengths, by other things that may be infinite. This is due to the fact that in floating point arithmetic, $\frac{\infty}{\infty}=\mathtt{NaN}$.} due to the need to gracefully handle infinite bounds (shown in \cref{fig:pickdef}) and avoid stability issues due to floating point arithmetic. Instead of allocating some fraction of the range to the left subsynthesis, and the rest to the right subsynthesis, the final algorithm initially allocates the full length to left, and none to the right. Then, some random amount of length (explicitly not a fraction) is transferred to it. This both avoids floating point division, and allows infinite bounds to be cleanly handled. We show a full implementation in \cref{fig:partition_add}.

\begin{figure}
\begin{minted}[breaklines, texcomments, linenos]{python}
# The bounds here are represented as a single array with either 2 or 4 points according to if the under bound is empty or not. The points are ordered in the intuitive order: for example, the double bound ${\text{«}}[u_l, u_h],\allowbreak[o_l, o_h]{\text{»}}$ is represented as $[o_l,\allowbreak u_l,\allowbreak u_h,\allowbreak o_h]$
def partition_add(vals):
    r1 = vals.copy() # One starts off full
    r2 = [0.0] * len(vals) # and the other empty
    currmag = 0.0 # How much length has been transfered
    for n in range(1, len(vals)):
        maxmag = diff(vals[n - 1], vals[n]) # Maximum transfer
        # Each subsequent point must carry the transfers of previous points, so we accumulate them in currmag
        currmag += pick(0.0, maxmag)
        r1[n] -= currmag
        r2[n] += currmag
    # Finally: offset the ranges in opposite directions
    perturbation = pick(-math.inf, math.inf)
    for i in range(len(r1)):
        r1[i] += perturbation
    for i in range(len(r2)):
        r2[i] -= perturbation
    return r1, r2
def diff(a, b):
    # Calculates the maximum length that can be transfered from a bound with endpoints $a$ and $b$
    if math.isinf(a) or math.isinf(b):
        return math.inf
    else:
        return b - a
\end{minted}
\caption{Implementation of the \textit{partition\_add} algorithm.}
\label{fig:partition_add}
\end{figure}

\begin{figure}
  \begin{equation*}
  \begin{split}
    \mathit{pick} [-\infty, +\infty]  & = \mathit{Normal}(0, 1) \\
    \mathit{pick} [l, +\infty]  & = l + \mathit{abs}(\mathit{Normal}(0, 1)) \\
    \mathit{pick} [-\infty, h]  & = h - \mathit{abs}(\mathit{Normal}(0, 1)) \\
    \mathit{pick} [l, h]  & = \mathit{Uniform}(l, h)
  \end{split}
    \end{equation*}
    \caption{Definition of pick. The exact distributions used here are an implementation detail. The most important aspect is that the supports match the requested range.}
    \label{fig:pickdef}
\end{figure}


\begin{figure}[htb]
  \begin{minipage}{0.5\textwidth}%
  \begin{minted}[escapeinside=||]{haskell}
|\tikzmark{exampleleft1}$\mathrlap{\mathrlap{\Circled{2\phantom{\mathclap{b=1}}}}\Circled{\phantom{2}\enspace\,b=1}\ {\text{«}}{\emptyset}, [-2.0, -1.0]{\text{»}}}$\hphantom{XXXXXXXXXXXXXXXXXXXXXXXXXXXXXXXX}|let Uniform(-1.7, -1.3) in| \Circled{3}|
|\tikzmark{exampleleft2}$\mathrlap{\mathrlap{\Circled{5\phantom{\mathclap{b=1}}}}\Circled{\phantom{5}\enspace\,b=1}\ {\text{«}}{\emptyset}, [0.0, 1.0]{\text{»}}}$\hphantom{XXXXXXXXXXXXXXXXXXXXXXXXXXXXXXXX}|let Uniform(0.0, 1.0) in| \Circled{6}|
|\tikzmark{exampleleft3}$\mathrlap{\mathrlap{\Circled{4\phantom{\mathclap{b=2}}}}\Circled{\phantom{4}\enspace\,b=2}\ {\text{«}}{\emptyset}, [0.0, 1.0]{\text{»}}}$\hphantom{XXXXXXXXXXXXXXXXXXXXXXXXXXXXXXXX}|let Beta(0.5, v2) in| \Circled{7}|
|\tikzmark{exampleleft4}$\mathrlap{\mathrlap{\Circled{1\phantom{\mathclap{b=4}}}}\Circled{\phantom{1}\enspace\,b=4}\ {\text{«}}[-1.0, 0.0], [-2.0, 1.0]{\text{»}}}$\hphantom{XXXXXXXXXXXXXXXXXXXXXXXXXXXXXXXX}|Uniform(v1, v3)| \Circled{8}|
\end{minted}
\begin{tikzpicture}[overlay, remember picture]
\draw[->]([shift={(0,1.0ex)}]pic cs:exampleleft4) .. controls ++(-1.5,0) and ++(-1.5,0) .. ([shift={(0,-0.1ex)}]pic cs:exampleleft1);
\draw[->]([shift={(0,1.0ex)}]pic cs:exampleleft4) .. controls ++(-0.5,0) and ++(-0.5,0) .. ([shift={(0,-0.1ex)}]pic cs:exampleleft3);
\draw[->]([shift={(0,1.0ex)}]pic cs:exampleleft3) .. controls ++(-0.5,0) and ++(-0.5,0) .. ([shift={(0,-0.1ex)}]pic cs:exampleleft2);
\end{tikzpicture}
  \end{minipage}
  \caption{Example of a synthesis with an empty context $\Delta$, type ${\text{«}}[-1.0, 0.0], [-2.0, 1.0]{\text{»}}$ and budget 4. From left to right: The arrows indicate that a line initiated the synthesis (and thus, decided the type of another line). The numbers circled represent the order of the steps taken. Those on the left, indicate when synthesis of that specific line started. Immediately to the right of those circles, the allocated budget ($b$) is shown. Further right of those numbers, we have the type that was requested to be synthesized for that particular line. Even further right, we show the synthesized program. Finally, the rightmost circles indicate when synthesis of that specific line was finished.}
  \label{fig:synthexample}
\end{figure}
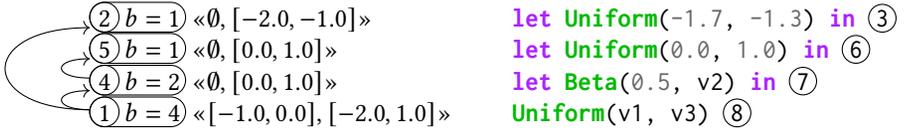

Let us now examine the synthesis example in \cref{fig:synthexample}:
\begin{enumerate}
  \item[\Circled{1}] Our synthesis begins with a target type of ${\text{«}}[-1.0, 0.0], [-2.0, 1.0]{\text{»}}$ and a budget of 4. A suitable distribution, in this case, the uniform distribution, is picked. Each node consumes 1 unit of budget, so 3 units are left to allocate among children (in this example, 1 unit is allocated to the left, and the 2 to the right). In addition, children are allocated types, (concretely, ${\text{«}}{\emptyset}, [-2.0, -1.0]{\text{»}}$ is assigned to the left, and ${\text{«}}{\emptyset}, [-2.0, -1.0]{\text{»}}$ is assigned to the right).
  \item[\Circled{2}] The first subtree gets the type of ${\text{«}}{\emptyset}, [-2.0, -1.0]{\text{»}}$ and the budget of 1. It consumes one unit of budget for itself, leaving 0 to allocate. With no budget left, this distribution can only use either constants or variables in the context. At this point, there are no variables in the context, and as such, terminal constants are synthesized.
  \item[\Circled{3}] Synthesis of this node is finished. The type ${\text{«}}{\emptyset}, [-2.0, -1.0]{\text{»}}$ is appended to the context, and is available for any further synthesized subtree to use as a variable.
  \item[\Circled{4}\Circled{5}\Circled{6}\Circled{7}] Having synthesized the left subtree of \Circled{1}, we must now synthesize the right subtree. This process follows in the same vein as the prior steps.
  \item[\Circled{8}] Finally, with all subtrees complete, our root node can be returned.
\end{enumerate}


\subsection{Soundness}
\label{subsec:soundness}

\paragraph{Soundness Theorem} All synthesized expressions have the type from which they were synthesized. I.e.,

 $\forall e, \synths{\Delta}{\Delta'}{{\text{«}}d{\text{»}}}{b}{e} \implies \Gamma \vdash e : {\text{«}}d{\text{»}}$, with $\Gamma \approx \Delta$.

 \paragraph{Proof} Proof is by induction on the synthesis rules:

\begin{itemize}%
 \item For the literal rule, the direct type of the synthesized literal $v$ is ${\text{«}}v,v{\text{»}}$, by \texttt{Type-Literal}. By \texttt{Type-Subtyping}, the singleton interval is the same as the requested underapproximation, and is contained in the overapproximation, given by the well-formedness of the requested type.
 \item For the range rule, the proof is the same, but the direct singleton type is as underapproximation of the provided range, because the value was sampled from a uniform with those same bounds.
 \item For the normal rule, the synthesis is symmetric to the type-checking rule, so all type-checking premises come from the synthesis requirements. This is similar to other distributions.
 \item For the uniform rule with an empty underapproximation (\texttt{Synth-Uniform-Loose}), the premises have half of the overapproximation interval (plus or minus the perturbation), which summed together are exactly the requested overapproximation. The underapproximations are always empty, thus being irrelevant in the proof.
 \item For the uniform rule with a non-empty underapproximation (\texttt{Synth-Uniform-Strict}), the proof relies on the generated ranges guaranteeing that the lower parameter cannot be higher than the lowest value of the underapproximation, and the higher parameter cannot be lower the highest value of the overapproximation. With these requirements, the coverage of the underapproximation is also guaranteed.
 \item For the add rule, the main proof centers on the (\texttt{partition\_add}) function~(\cref{fig:partition_add}). Throughout the function, the invariant that the sums of r1 and r2 is maintained throughout the whole function, except for two specific lines: First, in line 10, after which the invariant is immediately restored on line 11. And subsequently on the loop in line 14, whereupon the invariant is restored by the loop in line 16.
\end{itemize}

\section{Evaluation}
\label{subsec:results}

Our evaluation takes two fronts: Firstly, we compare with syntax-guided random search, which can generate complex trees but is rather simple and data-blind. Then, we compare with DaPPer~\cite{blogsynthesis}, a more advanced data-guided tool.

\paragraph{Dataset Selection}
The PSI~\cite{psippl} project has one of the most extensive and diverse test suites we could find. In addition, the availability of the ground truth programs facilitates both sampling for however many points we need, and filtering for only the supported features. Therefore, this test suit formed the basis for our benchmark dataset.

We collect a number of probabilistic programs from the PSI test set and filter those that use distributions or features that are supported by \lang{}, resulting in 105 programs. We then sample 10000 points from each of these programs. Each set of points forms a single test in our test set.

\paragraph{Comparison with Random Search}\label{comp1}
The evaluation consists of the following: given a time (60 seconds) and node budget, find the program that best fits a given set of samples. The fitness quality is evaluated using the Goodness of Fit test described in Bartoszynski et al.~\cite{triangletest}. Goodness of Fit tests attempt to quantify the similarity of two empirical distributions, or, in other words, how likely two sets of samples are to come from the same distribution.

We compare our synthesis with a type-agnostic syntax-guided random search. In our experiments, each test was instantiated with 20 different seeds, both for our method and random search. This results in a total of 40 runs per test.

\paragraph{Comparison with DaPPer}\label{dappercomp} We take advantage of the programs collected from \cref{comp1} to compare with DaPPer~\cite{blogsynthesis}. We use the system as described and published. As this tool's artifact does not allow for time limits, we run it for a set number of iterations (250, the default). We then evaluate our tool on the same problem for the amount of time that DaPPer took for those iterations. As such, both tools run for the same amount of time for each individual test.

\begin{figure}[htbp]
  \centering
  \includegraphics[width=0.8\linewidth]{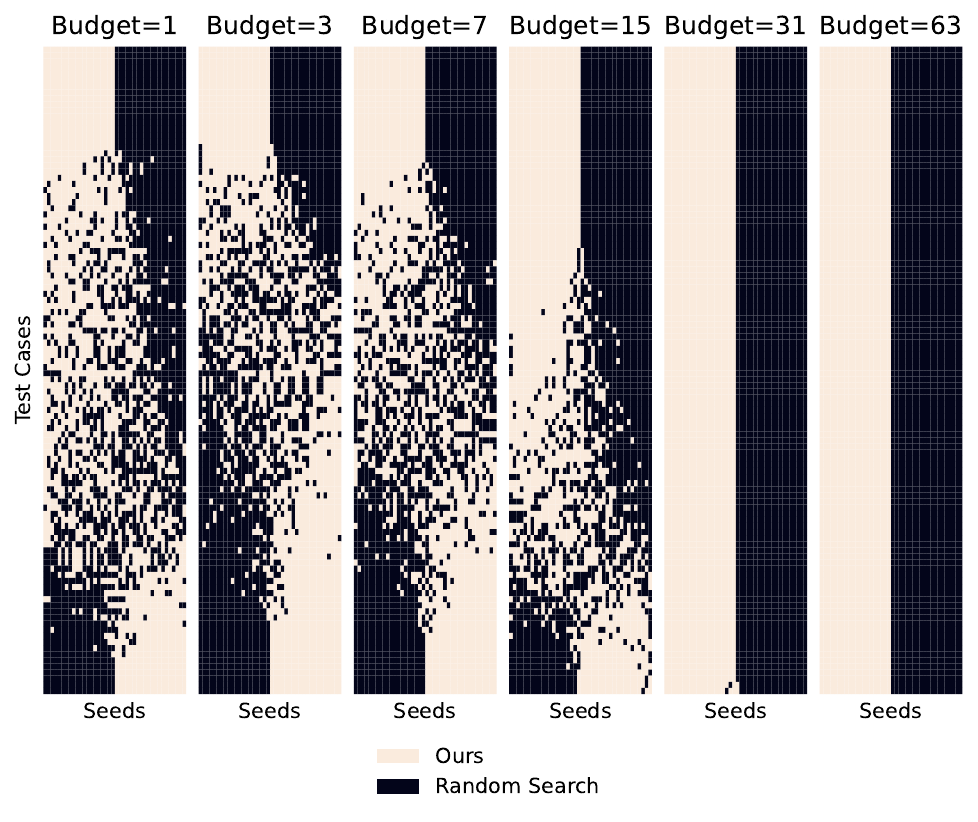}
  \caption{Sorted performance of our approach (tan) and random search (black) from best (left) to worse (right) fitness. Each test case (row) contains 40 runs (20 from each approach). Rows are ordered to maximize interpretability, with no other meaning. As the budget increases, our approach achieves better fitness in more examples.}
  \label{fig:results}
\end{figure}

\begin{figure}[htbp]
  \centering
  \includegraphics[width=0.8\linewidth]{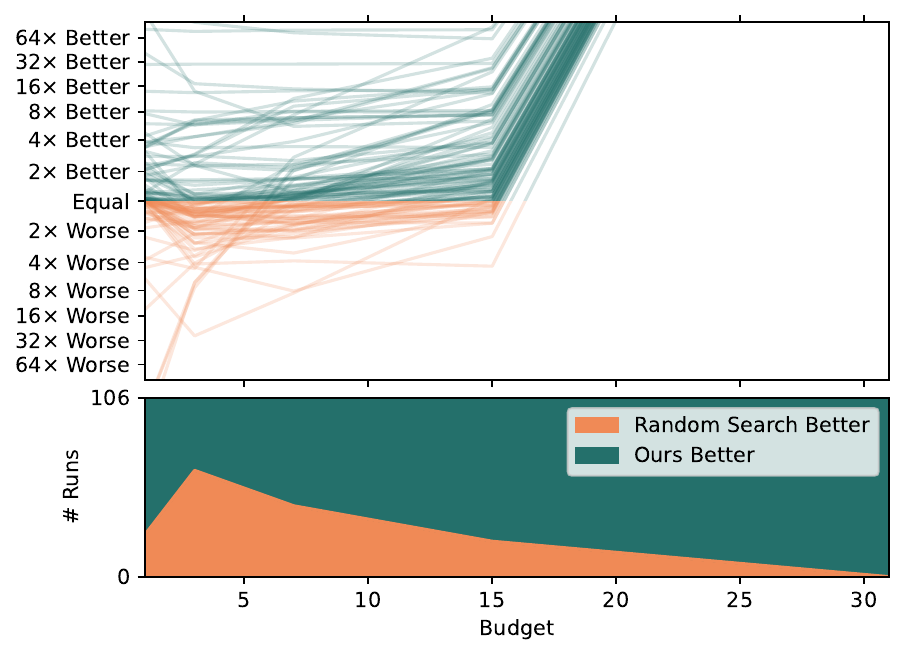}
  \caption{Comparison of fitness between our method and random search. The top plot shows a relative comparison: each line represents a specific test, with the Y axis representing how much better or worse it is. The bottom figure depicts the number of instances our algorithm did better, or worse, than random search.}
  \label{fig:relativeplot}
\end{figure}

\begin{figure}[htbp]
  \centering
  \includegraphics[width=0.5\linewidth]{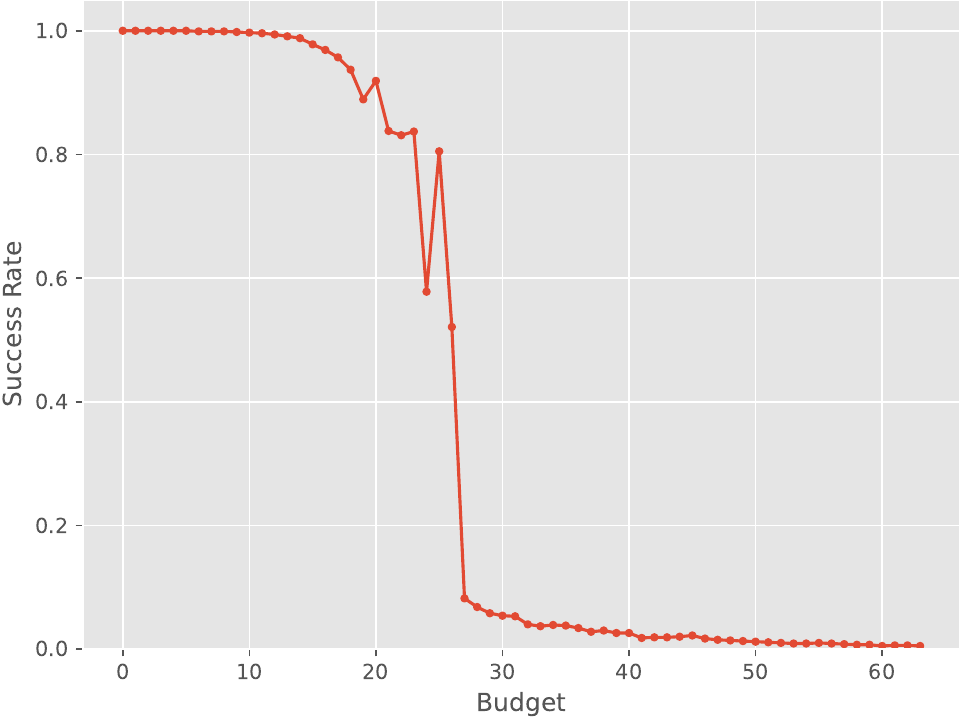}
  \caption{Success rate of random search in synthesizing a valid individual, as the budget increases. By observing this graph, we can observe the break-point at ($\mathit{budget}\approx26$).}
  \label{fig:succplot}
\end{figure}

\paragraph{Results}

\Cref{fig:results} shows a relative comparison between our method and random search. Each test (row) shows the 40 runs (ours is colored tan, while random search is colored black) sorted from best (left) to worst (right) fitness. Tests are clustered vertically without no other special meaning. It is visible that as the budget increases, our approach performs better until it is consistently better in all cases.

Additionally, \cref{fig:relativeplot} complements this analysis by quantifying how much better each test was in our approach (top) and by identifying how many runs performed better with our approach versus using random search. We can take the same conclusions as before: as the budget increases, not only our approach performs better in more runs, the value of fitness is also much better.

From these results, we can see that as the node budget increases, so does the advantage of our algorithm over random search. While at smaller budget values, we are just competitive, the improvement is already evident with $\mathit{budget}=15$. With $\mathit{budget}=31$, our algorithm eclipses random search in all but two test cases. To validate the reason for such increase, we analyze the success rate of random search in \cref*{fig:succplot}. As the synthesis of any single node can be seen as independent, each additional node synthesized carries with a new chance to fail. As such, our type-guided approach does not fail, always having some useful fitness value.

\begin{figure}[htbp]
  \centering
  \includegraphics[width=0.5\linewidth]{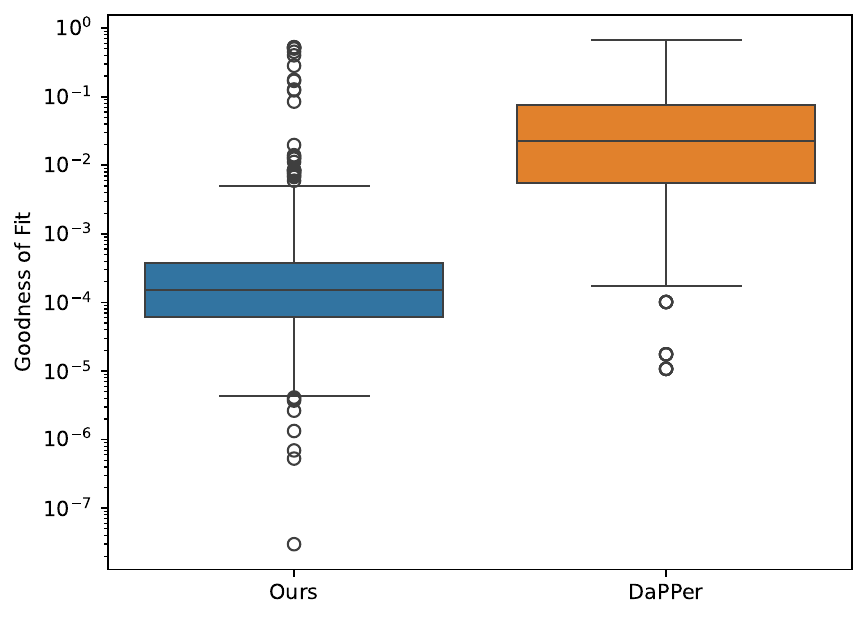}
  \caption{Box plot of the overall fitnesses obtained during evaluation in comparison with DaPPer~\cite{blogsynthesis}. Lower is better.}
  \label{fig:chasinsplot}
\end{figure}

Focusing now at the comparison with DaPPer in \cref{fig:chasinsplot}, our approach achieves models that better suite the data than DaPPer, with a difference in two orders of magnitude on the means of the distribution of the whole dataset.

While DaPPer is able to infer the structure from observed data, it requires variables relevant to the structure to be observed and present in the dataset (or, in other words, not hidden). In the benchmark we have used, that is not available, and neither it is in many real world problems~\cite{DBLP:journals/jstatsoft/MerkleFUG21}. Because of this,  DaPPer is unable to infer any kind of structure. Under this setting, DaPPer is only able to produce expressions consisting of a single distribution parameterized by constants.
In contrast, our method makes no assumptions about the non-existence of hidden variables. For these common scenarios, our approach finds better fitting solutions.

\section{Related Work}

In this section, we discuss existing techniques for the synthesis of probabilistic programs, highlighting their limitations. These limitations justify our proposed approach for Stochastic Symbolic Regression.

\subsection{Synthesis of Probabilistic Programs}

Regarding synthesis of probabilistic programs, Chasins et al.~\cite{blogsynthesis} describe a synthesis technique based on dependency analysis and simulated annealing. However, one key limitation of their approach is that it is unable to model hidden variables when they are not present in the dataset. Furthermore, the simulated annealing only optimizes parameters, not exploring different structures. Our synthesis algorithm can synthesize hidden variables, even without them being present in the dataset, and is able to achieve better models (approximately two orders of magnitude, on average) within the same time budget (\cref{dappercomp}).

Another synthesis technique is proposed in Saad et al.~\cite{baysynth}.  However, this method requires a model of models, i.e a sketch (or a grammar), and a prior for this model of models. Most importantly, it also requires the likelihood of the model of models to be tractable (which excludes some models). In addition, this likelihood calculation is sketch specific (thus, requiring expert knowledge to encode). Unfortunately, it was not possible to produce a comparison with this tool due to the artifact having suffered bit-rot\footnote{Even after pinning dependencies to their last Python 2 compatible versions, some had been removed from the Python repositories.}.

Finally, another direction is proposed by Andriushchenko et al.~\cite{DBLP:conf/tacas/Andriushchenko021, DBLP:conf/cav/AndriushchenkoC21}, wherein Markov Chains are used to induce probabilistic programs. However, this technique can only reason about events with a finite number of states (as such, it excludes real-valued events). In addition, similarly to Saad et al.~\cite{baysynth}, it requires a sketch to be provided.

These works are not adequate for the purpose of Stochastic Symbolic Regression, either by being unable to model hidden variables, or by requiring non-\textit{evidence} input.

ADEV~\cite{adev} extends traditional forward mode automatic diferentiation to better support probabilistic programming languages. It employs a kit of techniques described in the literature. It effectively transforms a distribution into a distribution of derivatives. However, while it can be used as efficient tool to fine-tune models from data, it does not attempt to infer model structure.


\subsection{Synthesis of Dependently Typed Programs}

Two works stand out regarding synthesis of refinement typed programs. Firstly, in Osera et al.~\cite{typeexamplesynth}, where examples are pushed through the tree during synthesis, resulting in invalid programs being refuted earlier. This is then used to eliminate classes of invalid programs from the search, speeding up synthesis. Despite decreasing the impact of backtracking through early detection, backtracking still happens. Our approach does not suffer from such penalty.

Synquid~\cite{synquid} expands on the previous paper by working directly with refinement types instead of examples, and abducting them during synthesis to trim the search. These abducted refinement types are then discharged using an SMT solver.

However, both these are not directly applicable to our work. This is due to three factors. Firstly, synthesizing a probabilistic program that \textit{could} output some given samples is a trivial problem. If the samples are real numbers, then any distribution over $\mathbb{R}$ (for example, $\mathit{Normal}(0,1)$) could, with nonzero probability, produce them. And this highlights the second issue: for proper exploration of the search space of possible models, the synthesized individuals should be diverse. But current approaches focus on correctness and not on diversity. The usage of SMT solvers in particular is problematic, as the many optimizations and shortcuts implemented to improve performance, result in a limited diversity of individuals, obtaining very simple solutions. Finally, both of these approaches are multistaged, wherein programs are produced by (albeit in a trimmed down search space) trial and error, instead of a single-pass, non-backtracking top-down procedure like ours.

\section{Discussion and Future Work}

Existing solutions for automatically learning symbolic probabilistic programs from observations are limited in either performance (by performing backtracking) or by their expressive power (being limited to explore parameters, but not the full search space of program structures; or by only modelling discrete values).

Our approach addresses both limitations at the same time: we allow the search of all possible syntactic expressions, while limiting the search to those are semantically valid. This validation is guaranteed by our type system, that can be extended with more distributions as long as the supports and images can be defined, using the parameters of the distribution. We show how many common distributions in PPLs have this requirement.

Our empirical evaluation in the PSI benchmark shows that our type-guided synthesis achieves better performance than syntax-guided random search when programs can be complex in the number of nodes. As programs become bigger, the type errors accumulate at each step, making synthesis not practical. Our approach allows synthesis of safe probabilistic programs to scale to program sizes not previously supported, and by having better performance, to larger datasets.

We also show that our synthesis has better performance (finds solutions two orders of magnitude better in the same amount of time) than DaPPer in the PSI benchmark, the state-of-the-art synthesis approach. DaPPer is limited by requiring hidden variables to be observed in the dataset (which does not happen in this benchmark, nor in many real-world problems). Dapper also focuses the search on finding the best parameters for a relatively simple structure, while our approach explores the search space of many valid structures.

To conclude, we present evidence that our approach can scale to more complex problems, which allows practitioners to automatically derive models for complex problems with large datasets.

For future work, we intend to explore more complex control flow constructs, such as iteration, and evaluate how this synthesis could be adapted dynamically to each given problem.

\begin{acks}
  This work was supported by Fundação para a Ciência e Tecnologia (FCT) through the LASIGE Research Unit (ref. UIDB/00408/2020 and UIDP/00408/2020), PhD fellowships (SFRH/BD/137062/2018 UI/BD/151179/2021 SFRH/BD/151469/2021), CMU|Portugal CAMELOT project (LISBOA-01-0247-FEDER-045915), FCT Exploratory project RAP (EXPL/CCI-COM/1306/2021) and FCT Advanced Computing projects (CPCA/A1/395424/2021, CPCA/A1/5613/2020, CPCA/A2/6009/2020). We also thank the anonymous reviewers for their valuable feedback.
\end{acks}

\printbibliography

\end{document}